\documentclass[usenatbib,useAMS]{mn2e}

\usepackage[dvips]{graphicx}
\usepackage{dcolumn}
\usepackage{multirow}
\usepackage{times}

\newcommand{\mnras}{MNRAS}
\newcommand{\apj}{ApJ}
\newcommand{\aap}{A\&A}
\newcommand{\aj}{AJ}
\newcommand{\apjl}{ApJ}
\newcommand{\apjs}{ApJS}

\newcommand{\aaps}{A\&AS}
\newcommand{\araa}{ARA\&A}

\title{Global 8.4-GHz VLBI observations of JVAS B0218+357}
\author[A.~D.~Biggs et al.]{A.~D.~Biggs,$^{1,2}$\thanks{E-mail: biggs@jive.nl} 
O.~Wucknitz,$^{3,1}$\thanks{Present address: Universit\"{a}t Potsdam,
  Institut f\"{u}r Physik, Am Neuen Palais 10, 14469 Potsdam, Germany.}
R.~W. Porcas,$^4$ I.~W.~A. Browne,$^1$ N.~J.~Jackson,$^1$
\newauthor S.~Mao$^1$ and P.~N.~Wilkinson$^1$\\
$^1$University of Manchester, Jodrell Bank Observatory, Macclesfield, Cheshire 
SK11 9DL\\
$^2$Joint Institute for VLBI in Europe, Postbus 2, 7990 AA Dwingeloo,
The Netherlands\\
$^3$Hamburger Sternwarte, Universit\"{a}t Hamburg, Gojenbergsweg 112,
D-21029 Hamburg, Germany\\
$^4$Max-Planck-Institut f\"{u}r Radioastronomie, Auf dem H\"{u}gel 69,
D53121, Bonn, Germany\\}
\begin{document}

\maketitle
\begin{abstract}

In this paper we present new observations of the gravitational lens
system JVAS B0218+357 made with a global VLBI network at a frequency of 
8.4~GHz. Our maps have an rms noise of 30~$\mu$Jy~beam$^{-1}$ and with
these we have been able to image much of the extended structure of the
radio jet in both the A and B images at high resolution ($\sim$1~mas).
The main use of these maps will be to enable us to further constrain
the lens model for the purposes of $H_0$ determination.
We are able to identify several sub-components common to both images
with the expected parity reversal, including one which we identify as a
counter-jet. We have not been successful in detecting either the core
of the lensing galaxy or a third image. Using a model of the lensing
galaxy we have back-projected both of the images to the source plane
and find that they agree well. However, there are small, but
significant, differences which we suggest may arise from multi-path
scattering in the ISM of the lensing galaxy. We also find an exponent
of the radial mass distribution of $\beta\approx1.04$, in agreement
with lens modelling of published 15-GHz VLBI data. Polarisation maps of
each image are presented which show that the distributions of
polarisation across images A and B are different. We suggest that this
results from Faraday rotation and associated depolarisation in the
lensing galaxy.

\end{abstract}


\section{Introduction}

Few gravitational lens systems are as well studied as JVAS B0218+357
\citep{patnaik93}, perhaps the best example of a lens system
for which the method of \citet{refsdal64} can be used to determine the
Hubble parameter, $H_0$. Unlike many other methods, this is done in a
single step and involves well understood, and relatively simple,
astrophysics. We point out that there remain significant uncertainties
in $H_0$ as determined by traditional methods \citep[e.g.][]{shanks01}.
For example, the use of Cepheid-calibrated distances to determine the
distance to galaxies hosting Type Ia supernovae has resulted in final
values of $H_0$ that do not agree at the $1\sigma$ ($\sim$10~per~cent)
level \citep{parodi00,freedman01}. This is despite a large overlap 
between the samples of SNIa.

Whilst the gravitational lens route to $H_0$ can be a particularly
``clean'' one, and despite many lens systems now having measured time
delays, almost all suffer in one way or another from effects that
significantly increase the uncertainty in the final determination of
$H_0$. In most cases the main source of uncertainty is in the modelling
of the gravitational potential responsible for the image splitting and
distortion. A lack of observational data to constrain the lens model
and multiple-galaxy lenses are both significant factors in this regard
\citep[see e.g.][]{schechter00}. A list of the time delays measured to
date can be found in \citet*{courbin02}.

With JVAS B0218+357 on the other hand, the potential exists to reduce
the uncertainty in the lens model to a level comparable to that in
the time delay; this currently stands at 3~per~cent ($\tau = 10.5\pm0.4$~d;
Biggs et al., 1999). Other authors find a value for the time delay
consistent with this \citep{cohen00}. This reduction in model
uncertainty is possible due to the large number of observational model
constraints available from multi-frequency, multi-resolution
imaging. Both images (A and B) of the $z=0.96$ background quasar/BL~Lac
are easily resolved with VLBI into two subcomponents
\citep*[e.g.][]{patnaik95,kemball01} which constrain the radial mass
profile to be close to isothermal. The lens systems with the largest
numbers of model constraints are those that contain Einstein rings,
such as B0218+357. The resolved arcsec-scale structure of the ring
probes the lens potential along multiple lines of sight and along
all azimuthal position angles relative to the lens centre. The ring 
(see Fig.~\ref{vlamap}) has been mapped at high resolution using
combined MERLIN and VLA data at 5~GHz \citep{biggs01} and the model
constrained using a version of the LensClean algorithm
\citep{kochanek92}; preliminary results can be found in 
\citet{wucknitz01}. Finally, the deflecting mass is concentrated in a
single, isolated galaxy and so the effect of tidal shear due to nearby
structure is very small ($\sim$1~per~cent, Leh\'{a}r et al. 2000). This
results in a relatively uncomplicated mass model compared to other lens
systems. A VLA map of B0218+357 can be seen in Fig.~\ref{vlamap}.

The main stumbling block to date in measuring $H_0$ with JVAS
B0218+357 has been that, due predominantly to the small size of the
system (A-B separation of 334~mas), it has been difficult to measure
the position of the $z=0.6847$ lensing galaxy relative to the lensed
images accurately using optical ($HST$) data. For example, the
positions derived from two NICMOS 
observations differ by 46~mas \citep{lehar00} and this uncertainty
translates into a large uncertainty in $H_0$. With LensClean we
have estimated the galaxy position to a theoretical accuracy of a few
mas, but this relies to some extent on the lens model and so could be
biased.

\begin{figure*}
\begin{center}
\includegraphics[scale=0.35]{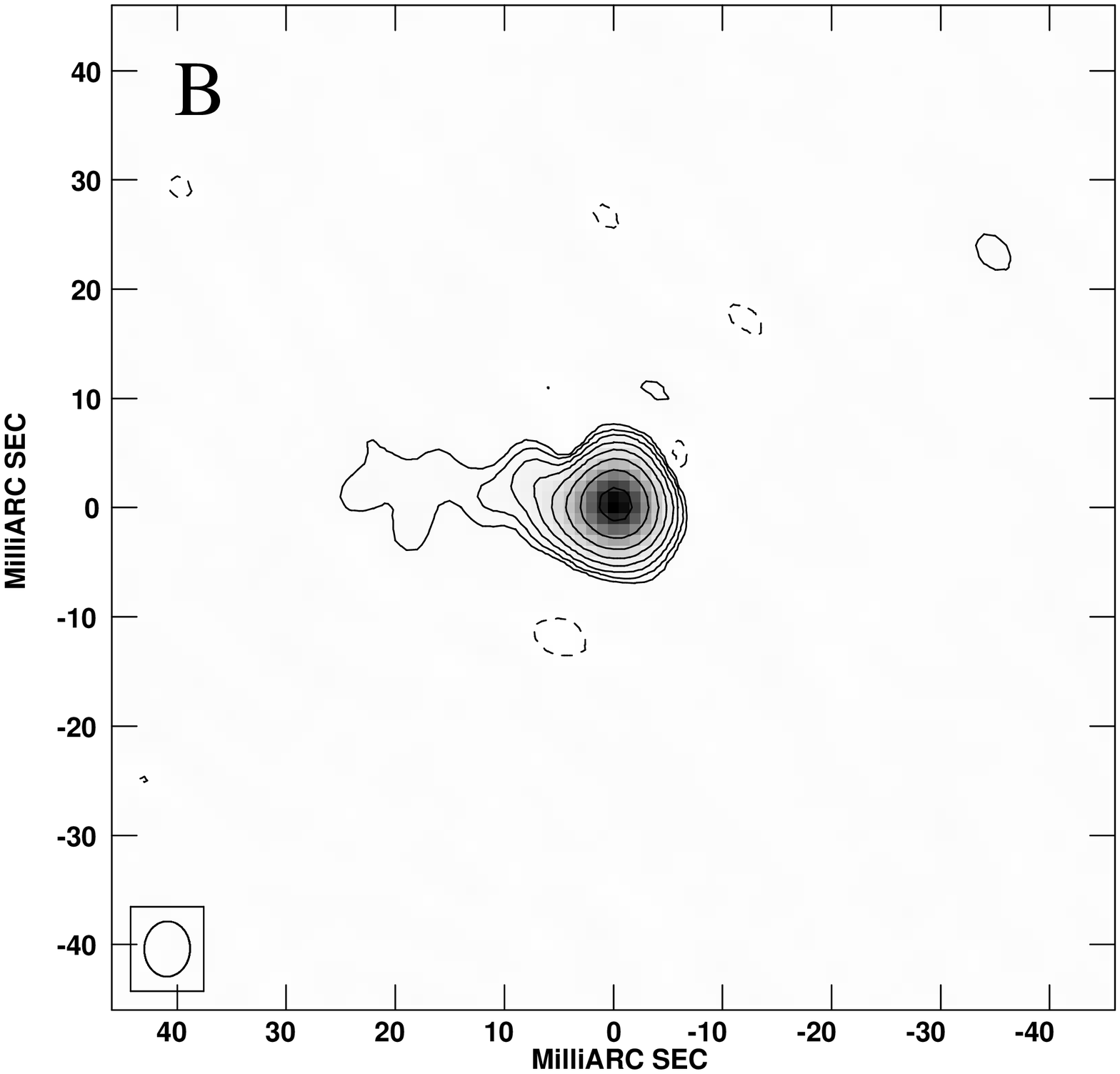}
\includegraphics[scale=0.35]{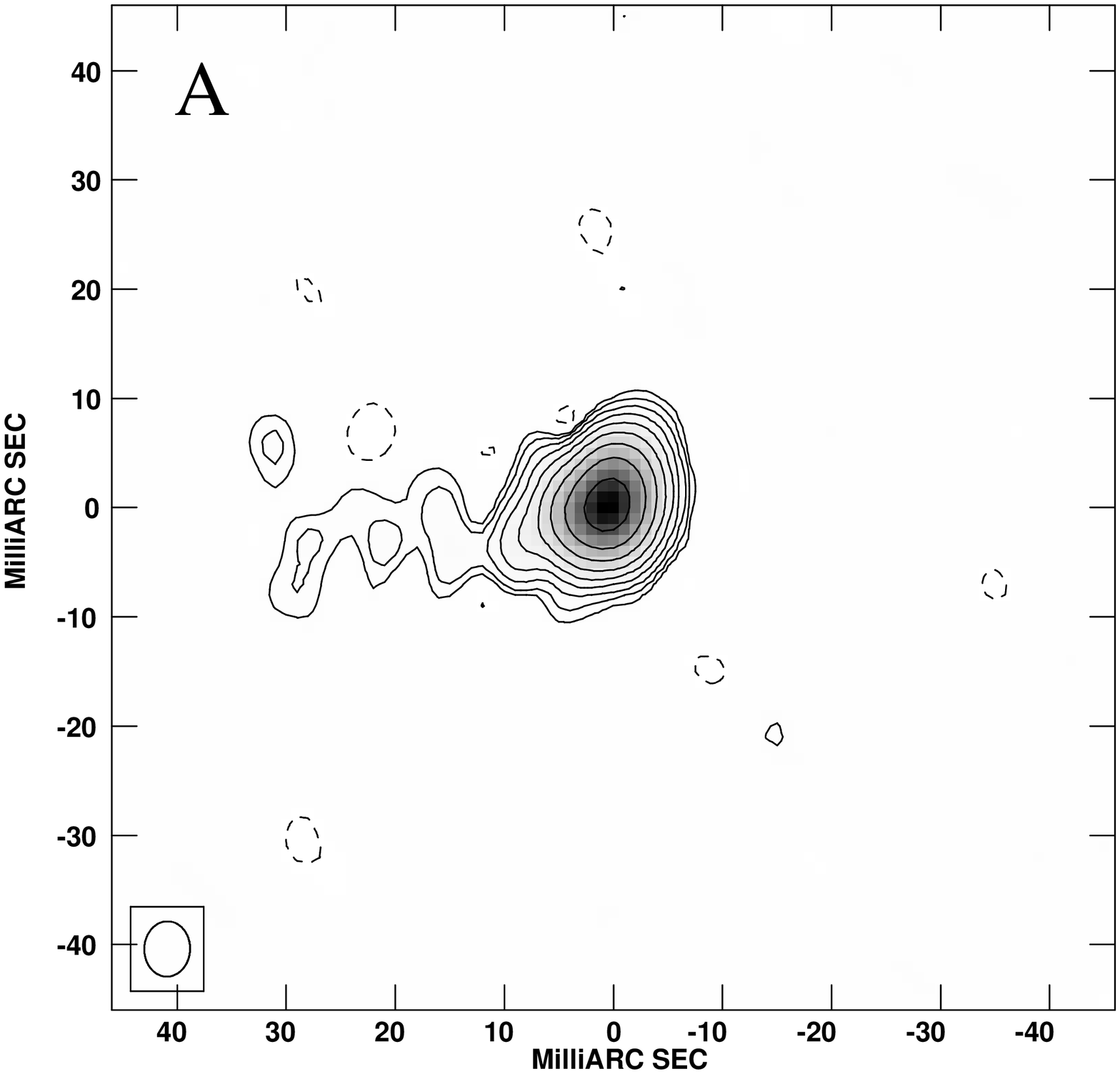}
\caption{CJ1 VLBI maps of JVAS B0218+357 at 5~GHz. Left: B, right: A. The
  restoring beam is shown in the bottom-left corner of each map and has
  a FWHM of $5 \times 4.2$~mas at a position angle of $-1$\fdg9. Contours
  are plotted at multiples ($-1$, 1, 2, 4, 8, 16, etc) of 3$\sigma$
  where $\sigma$ is the off-source rms noise in the map
  (330~$\mu$Jy~beam$^{-1}$). Both maps are plotted on the same angular
  scale.}
\label{cj}
\end{center}
\end{figure*}

In this paper we present new 8.4-GHz observations of JVAS B0218+357
taken with a global VLBI array that were designed to increase the
available model constraints by other means. With the great
sensitivity of a global array we hoped we might detect:
\begin{enumerate}
\item extended structure associated with the radio jet in both A and B.
  This is detected in 5-GHz data from the First Caltech-Jodrell Bank
  (CJ1) VLBI survey \citep{xu95} that we have re-analysed to produce
  maps (Fig.~\ref{cj}) that are superior to those published in
  \citeauthor{xu95}. With a resolution of $\sim$5~mas these new maps reveal
  a jet that extends out to 30~mas from the core in each image.
\item a third image, as detected in, e.g. APM~08279+5255
  \citep{ibata99,lewis02} and MG~1131+0456 \citep*{chen93,chen95}, and
\item the galaxy core. Emission from the lensing galaxy has been
  detected in several systems e.g. B0957+561 \citep{harvanek97} and
  CLASS B2045+265 \citep{fassnacht99}. A third radio component is
  detected in PMN~J1632-0033, but it is not known if this is a lensed
  image or the lens galaxy \citep{winn02}.
\end{enumerate}

The observations have also been used to investigate the mas-scale
polarisation structure of this system. Whilst the cores of most
quasars/BL~Lacs are polarised at a level of about 2--3~per~cent
\citep{saikia88}, images A and B of B0218+357 are both much more highly
polarised, reaching $\sim$10~per~cent at frequencies 8.4~GHz and above
\citep{patnaik93,biggs99}. Also, the polarisation position angles of
the images are not the same, in conflict with the expected behaviour of
a lens system which preserves the polarisation position angle on the
sky \citep{dyer92}. It is thought that Faraday rotation in the
magnetoionic medium of the lensing galaxy itself rotates the position
angles of the images by different amounts \citep{patnaik93}. Thus the
extended nature of the emission in images A and B allows the
interstellar medium (ISM) of a high-redshift galaxy to be studied on
parsec scales. Both images also depolarise at low frequencies.

\section{Observations and data reduction}

The goals of this experiment required a combination of high
sensitivity, high resolution and excellent $(u,v)$ coverage in order to
reliably image complex and faint structures. For this reason we used a
global VLBI array including the ten antennas of the VLBA, six from the
European VLBI Network (EVN), the VLA (Y) in phased-array mode and the
two 70-m antennas of the Deep Space Network (DSN) at Goldstone and
Robledo. Details of these antennas are given in Table~\ref{antennas}
with their diameters and system equivalent flux densities (SEFD).
We chose to observe at 8.4~GHz instead of at 5~GHz (where the surface
brightness sensitivity is better) due to the improved resolution and
because there is less chance that the lensed emission would be affected
by scattering; there is evidence that the image sizes in this system
may be enlarged by scattering at low frequencies (Biggs et al., in
preparation, although see \citet{porcas96b} for a 1.7-GHz map
that illustrates the large size increase). Scattering would
possibly reduce the surface brightness of the third image or the galaxy
core which is clearly undesirable when searching for such weak features.

\begin{table}
\begin{center}
\begin{minipage}{0.5\textwidth}
\renewcommand{\thefootnote}{\alph{footnote}}
\caption{Global VLBI array antennas and performance at
  8.4~GHz. Observations were conducted in dual circular polarisation
  at all antennas apart from Nt, On and Yb where only right circular
  polarisation was available.}
\begin{tabular}{lll} \hline
Antenna          & Diameter (m) & SEFD\footnote{System equivalent flux density} (Jy) \\ \hline
VLBA             & 25  & 307  \\
VLA (Y$_9$)\footnote{Inner nine antennas of VLA phased together} &
75\footnote{Diameter corresponding to summed geometric area of telescopes}  & 41  \\
Effelsberg (Eb)  & 100 & 20   \\
Medicina (Mc)    & 32  & 270  \\
Noto (Nt)\footnote{Only right circular polarisation was available}& 32  & 770  \\
Onsala (On)\footnotemark[4]      & 20  & 1630 \\
Westerbork (Wb)  & 94\footnotemark[3]  & 120  \\
Yebes (Yb)\footnotemark[4]       & 14  & 3300 \\
Robledo (Ro)     & 70  & 23   \\
Goldstone (Go)   & 70  & 20   \\ \hline
\end{tabular}
\label{antennas}
\end{minipage}
\end{center}
\end{table}

Due to the fact that the VLA was in its largest configuration (`A') at
the time of the observations, using the full complement of 27 antennas
as a single phased array was not possible as the synthesised beam would
have been significantly smaller than the separation between the two
images A and B. Therefore we 
were forced to use only the inner nine antennas of the array, Y$_9$, to
give a large enough beam. It should be noted that in order to calibrate
the absolute polarisation position angle, the strategy for which is
described later in this section, three short additional observations of
3C48 and 3C138 were added to the inner subarray schedule. These two
sources were {\em not} observed by any other antenna in the global
array. The remaining 18 antennas of the VLA observed as an outer
subarray operating in normal interferometric mode.

The observations took place during 11 November 2000 to 12 November 2000
over the time range 20:00--07:00 UT for the European antennas and
21:00--09:00 UT for those in the US. Both senses of polarisation were
recorded with the exception of three of the stations where only right
circular polarisation (RCP) was available (Nt, On and Yb).
Data were recorded in four 8~MHz sub-bands with one-bit sampling giving
a data rate of 128~Mbps. Of the 19 antennas scheduled (VLBA+EVN+DSN+Y),
only 17 took part in the observations; Yebes suffered from a broken
data formatter whilst the receiver at VLBA Owens Valley was
non-operational. The various calibration sources observed are shown in
Table~\ref{sources} along with their flux densities. The data were
correlated at the VLBA correlator in Socorro producing 16 0.5-MHz
channels per sub-band (IF) and using an averaging time of two
seconds. All polarisation pairs (RR, LL, RL and LR) were produced. 

\begin{table*}
\begin{center}
\begin{minipage}{3.5in}
\renewcommand{\thefootnote}{\alph{footnote}}
\caption{Approximate flux densities, hour angle coverage and
   role of the various calibration sources.}
\begin{tabular}{llll} \hline
Source     & Flux density (Jy) & \# Hour angles & Calibration role \\ \hline
B0234+285  & 3.0               & 29 & amplitude \\
3C84       & 19.0              & 8  & bandpass/L-R delay/D-term \\
0059+581   & 1.5               & 3  & polarisation position angle \\
0300+470   & 1.3               & 3  & polarisation position angle \\
3C48\footnote{VLA inner antennas only} & 3.2 & 2 & absolute polarisation position angle \\
3C138\footnotemark[1] & 2.4    & 1  & absolute polarisation position angle \\
\end{tabular}
\label{sources}
\end{minipage}
\end{center}
\end{table*}

Data reduction was performed using the NRAO {\sc aips}
package. Amplitude calibration was carried out using system 
temperatures ($T_{\mathrm{sys}}$) recorded during the observations scaled by
{\em a priori} measurements of the antenna gains. For the DSN antennas
no $T_{\mathrm{sys}}$ measurements were available, but values were estimated
from measurements of the total power. Delay residuals were calculated
using short segments of B0234+285 data 
and applied to all sources. This source was then fringe-fitted, solving
for delays, rates and phases for the entire dataset. Following this the
calibrated B0234+285 data were mapped in {\sc imagr} and
self-calibrated in {\sc calib}. The residual amplitude voltage gain
solutions found during the 
self-calibration were large for many antennas (up to a factor of two
for VLBA Pie Town) and so were subsequently
applied to all sources in order to correct the initial amplitude
calibration. Next the B0218+357 data were fringe-fitted, initially with
a single component located at the location of the brighter A component
relative to the correlation position, but later with a more realistic
model made from mapped and self-calibrated data. The delay difference
between the left and right polarisations was calibrated using a
single scan of 3C84 data and two baselines to the reference antenna
(VLBA North Liberty). The 3C84 data were also used to calibrate the
bandpasses of the antennas.

Next the various steps to calibrate the polarisation were
undertaken. Firstly, the leakage (or $D$-) terms were solved for using
the 3C84 data which had first been mapped and self-calibrated; 3C84 was
assumed to be unpolarised. Secondly the absolute position angle of the
polarisation was calibrated using the data from the inner subarray of
the VLA (Y$_9$) in normal 
interferometric mode\footnote{The antennas of the phased array also
produce the usual baseline combinations as a matter of course.}. As
mentioned earlier, the bright calibration sources 3C48 and 3C138 were
included in the Y$_9$ schedule and with these it was possible to
determine the true position angle of the source 0059+581. Once this
was known it was possible to calibrate the position angle of the global
array data by rotating the RCP$-$LCP phase difference until the 0059+581
polarisation position angle agreed with that found with the VLA data.

At this point the calibrated B0218+357 data were averaged in frequency
to produce one channel per IF and averaged in time over 10~s intervals.
The regions around images A and B were CLEANed using the {\sc imagr}
wide-field mapping task and the data self-calibrated in {\sc calib}.
Many cycles of mapping and self-calibration were undertaken, initially
only allowing the phases to vary, but later the amplitudes as well.
It was found that the sensitivity of the Westerbork (Wb) telescope was
changing by up to a factor of four, cycling over 20~min
timescales. Due to this, and the fact that the polarisation leakage 
terms were excessively large (up to 45~per~cent), the Wb data were
flagged and took no further part in the analysis. Much of the previous
calibration was then repeated to make sure that the Wb data was not
biasing the calibration for the other antennas. Data from the Noto and
Onsala antennas could not be included when making the Stokes $Q$ and
$U$ maps as neither was able to observe both right and left circular
polarisations.

Unlike the antennas that formed the global VLBI array
(VLBA+EVN+DSN+Y$_9$) the outer subarray of the VLA observed with a
separate schedule, nodding between B0218+357 (15 mins on source) and
0234+285 (2 mins on source) over a period of six hours. The data were
calibrated in {\sc aips} using 0234+285 as the phase, amplitude and
polarisation calibrator. The $D$-terms were solved for assuming a
point-source for 0234+285 (a good approximation to the polarised
intensity image) whilst simultaneously solving for the source
polarisation. The position angle of polarisation was again aligned
using the inner VLA subarray (Y$_9$) data. These data were then mapped
and self-calibrated, again in {\sc aips}.

\section{Results}

\subsection{Global VLBI maps (total intensity)}
\label{totintsec}

\begin{figure*}
\begin{center}
\includegraphics[scale=0.41]{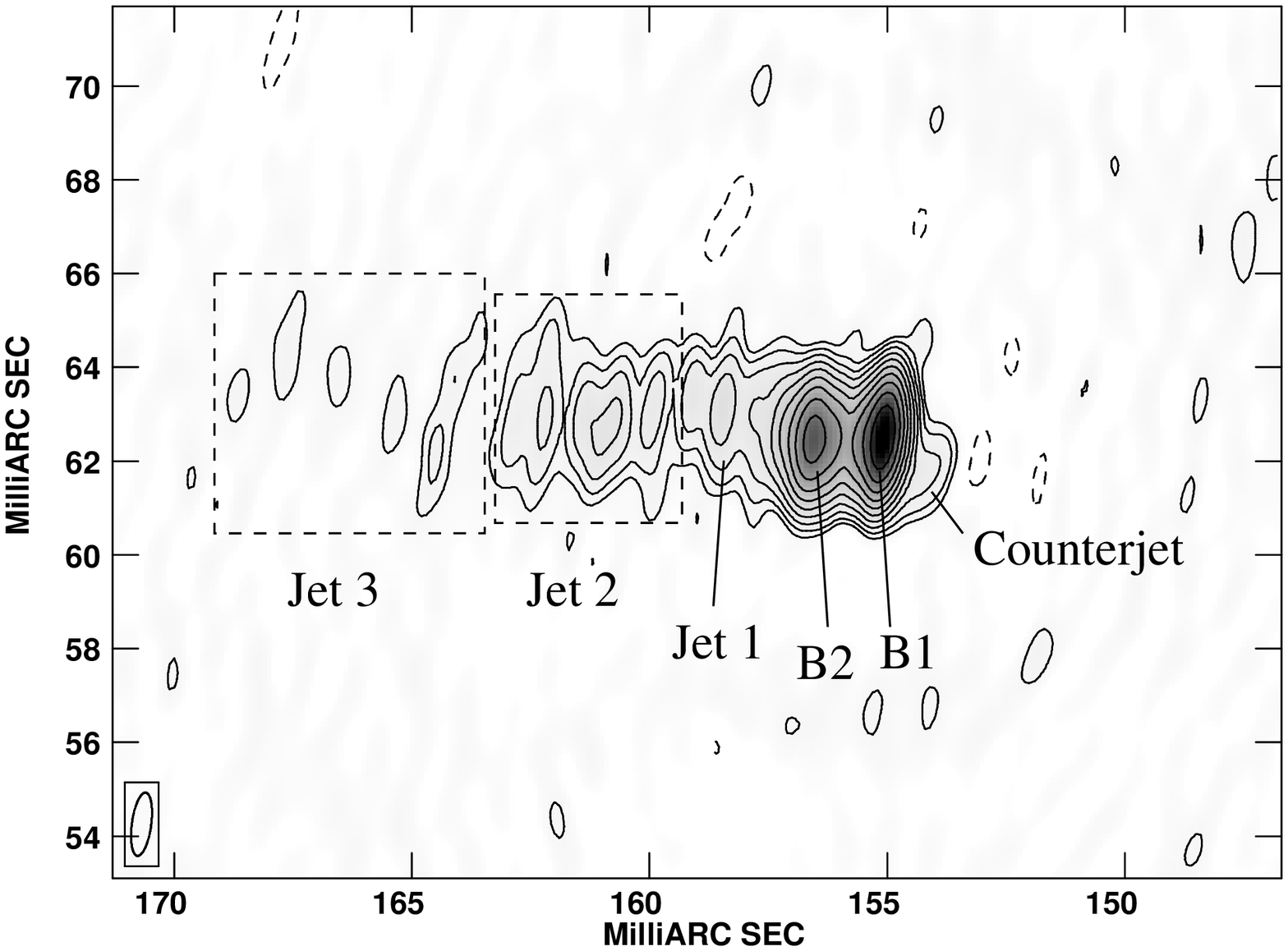}
\includegraphics[scale=0.41]{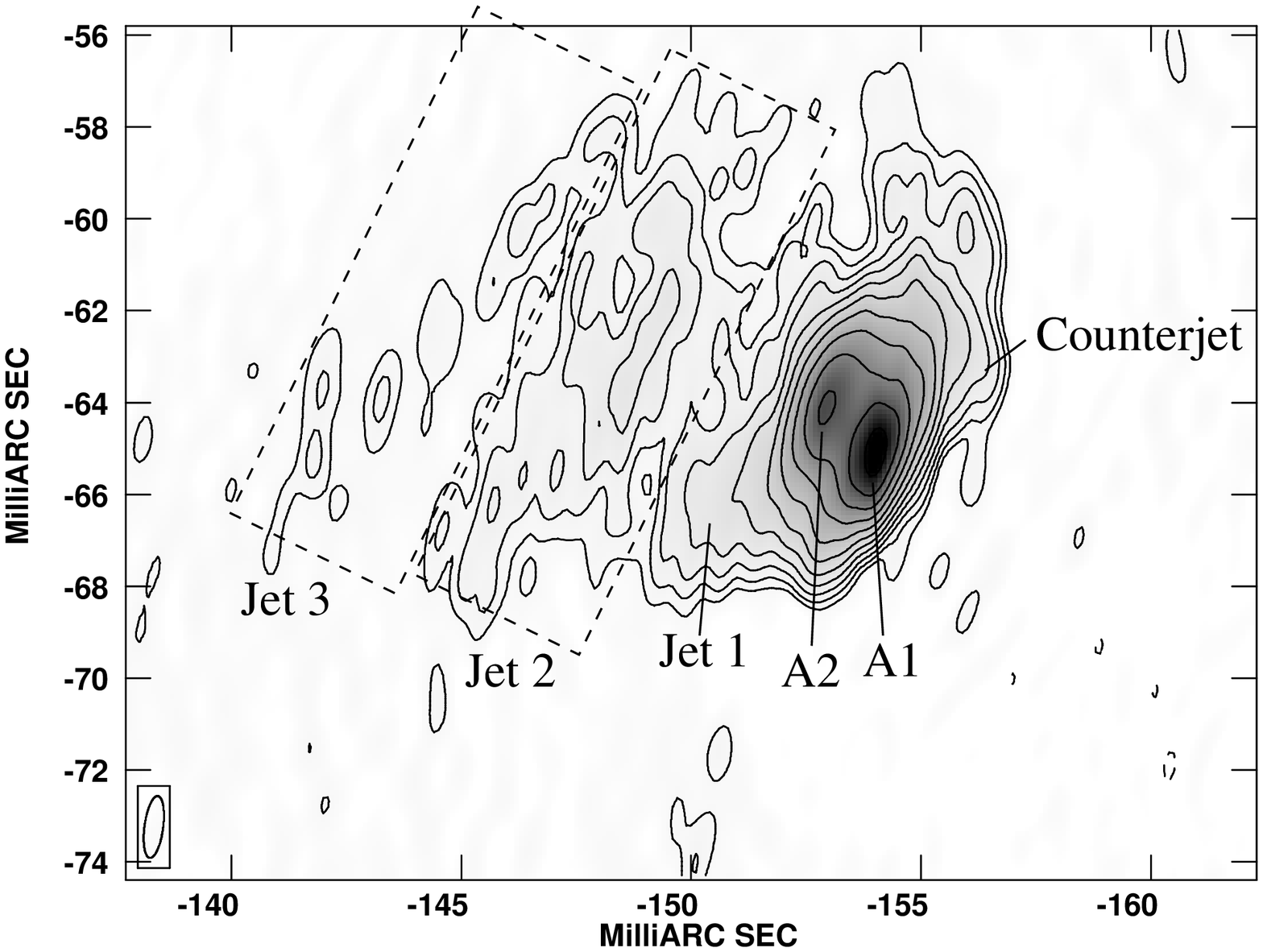}
\includegraphics[scale=0.43]{fig2d.ps}
\includegraphics[scale=0.43]{fig2c.ps}
\caption{Global VLBI maps of JVAS B0218+357 at 8.4~GHz. Left: B, right: A.
  Top: Total intensity (Stokes $I$). Contours are plotted at multiples
  ($-1$, 1, 2, 4, 8, 16, etc) of 3$\sigma$ where $\sigma$ is the
  off-source rms noise in the map (30~$\mu$Jy~beam$^{-1}$). The
  restoring beam is shown in the  
  bottom-left corner and has a FWHM of $1.36 \times 0.41$~mas 
  at a position angle of $-7$\fdg5. Bottom: Polarised intensity plotted
  on the same angular scale as the figures in the top row. Contours
  are also plotted at the same levels as in the top figures, but the
  restoring beam here has a FWHM of $1.27 \times 0.38$~mas 
  at a position angle of $-8$\fdg9.}
\label{global1}
\end{center}
\end{figure*}

The final VLBI maps are shown in Fig.~\ref{global1} and are made from
uniformly-weighted data in all cases. This results in an off-source rms
noise of 30~$\mu$Jy~beam$^{-1}$ (compared to a theoretical noise of
$\sim$10~$\mu$Jy~beam$^{-1}$) and a synthesised beam of $1.36
\times 0.41$~mas for the total intensity maps; these are shown in the
top two panels. Here the images are dominated by two bright
sub-components, with a separation of $\ga$1~mas in both A and B. These
were designated (A or B) 1 and 2 by \citet{patnaik95} and
observations at higher frequencies (15, 22 and 43~GHz) show the same
two sub-component structure \citep{porcas96a}. The
core is believed to be the most westerly sub-component (1) in each
case, based on its compactness relative to sub-component 2. Its
spectrum also turns over at a higher frequency, as expected for a more
compact component. As well as the two bright sub-components though,
we have also detected large quantities of extended emission, in both A
and B. This, however, looks quite different in each of the images, image
A being considerably more resolved than image B. It is exactly these
differences that can be used as constraints on the lens model as they
must be explicable, in the absence of effects such as scattering, by
the action of the lensing galaxy.

Interpreting image B is straightforward as it looks like a typical
core-jet radio source, with a knotty jet extending eastwards from the
bright core region. This is in line with what we would expect given the
new CJ maps of Fig.~\ref{cj}. However, in order to aid the eye in
comparing images A and B, the structures that we will now describe have
been labelled on the maps in Fig.~\ref{global1}. The jet in B appears to
bend, pointing in an approximate north-easterly direction immediately
after emerging from the core (Jet 1), before gently curving over the
next 5~mas to point in a more south-easterly direction (Jet 2). Finally
the jet changes direction again, returning to a north-easterly
trajectory before fading completely (Jet 3). Returning to the core
region, there is a small counter-component that appears to emerge from
the south-west of B1.

In image A sub-components 1 and 2 are, as already stated,
clearly detected. However, they appear to sit in a large structure that
is edge-brightened on its western side and extended to the
north and south. This has been observed before at this frequency
\citep{kemball01} as well as at 15~GHz \citep{patnaik95}, but the greater
sensitivity of our maps makes the effect much more pronounced. The
majority of this structure, as has been noted by previous authors, is
in fact emission from the core region that has been tangentially
stretched along a position angle $\sim-30^{\circ}$ (perpendicular to
the A-lens galaxy separation) as part of the lensing process. Lens
models predict just such a stretching of the brighter image in
two-image lens systems, but it is 
not usually as resolved as that seen in B0218+357. We can quantify the
magnitude of the stretching by noticing that the separations A1-A2 and
B1-B2 are approximately equal ($1.45\pm0.02$~mas). From observations
with MERLIN and the VLA the flux density ratio between A and B at
centimetre wavelengths is known to lie between three and four
\citep{patnaik93,biggs99,cohen00}. As this flux density ratio is due to an
increase in the size of image A relative to B then image A must be
stretched by a factor 3--4 relative to B in the observed direction. A
similar conclusion was reached by \citet{patnaik95}.

The other features identified in image B are also clearly visible in
image A and with properties consistent with gravitational lensing. The
jet emerges from the core region in A and is parity reversed as
expected -- whilst Jet 1 in B initially lies above a line joining B1
and B2, in A it lies below a line joining A1 and A2. The tangential
stretching displaces this jet feature away from the 1-2 separation
vector compared to B. Proceeding eastwards the jet emission remains 
tangentially stretched (the Jet~2 region is seen in the CJ1 map
(Fig.~\ref{cj}) as a bright extension to the east of the core) before
fading as in B into discrete blobs. Note, 
however, the opposite direction of the very faint emission at the far
end of the jet (Jet 3). In A this lies along a south-east line compared to
north-east in B. The counter-component in A is also visible, but is
located to the north of the A1-A2 separation, as opposed to south of
the B1-B2 separation vector.

\begin{figure*}
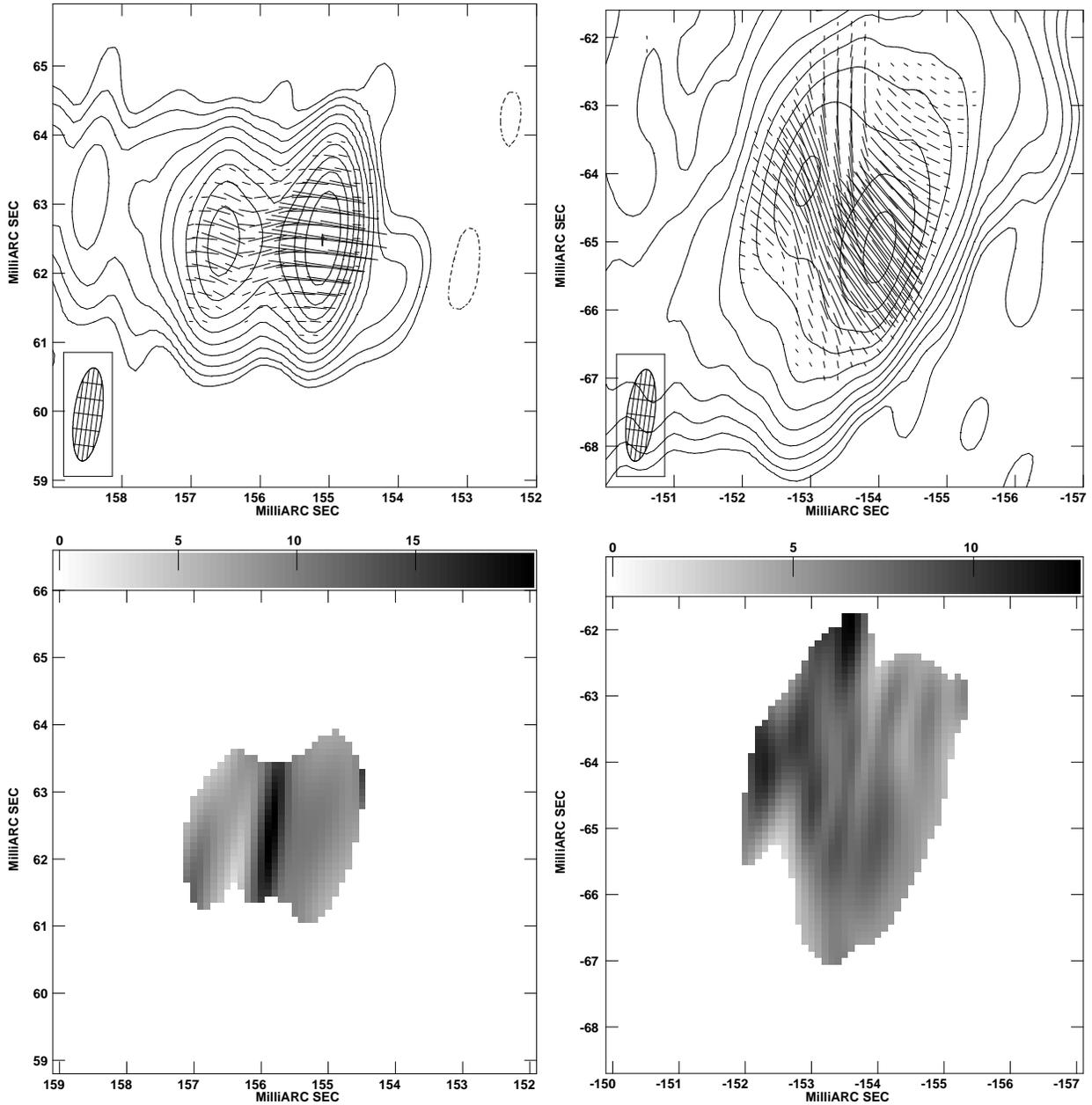

\begin{center}
\includegraphics[scale=0.43]{fig3b.ps}
\includegraphics[scale=0.43]{fig3a.ps}
\includegraphics[scale=0.43]{fig3d.ps}
\includegraphics[scale=0.43]{fig3c.ps}
\caption{Global VLBI maps of JVAS B0218+357 at 8.4~GHz. Left: B, right:
  A. Top: Close up of the total intensity maps (contours) of
  Fig.~\ref{global1} with polarised intensity E-vectors overlaid. A 
  vector of length 1~mas equals 5~mJy~beam$^{-1}$. Bottom: Percentage
  polarisation greyscales plotted on the same angular scale as the
  figures in the top row. For each image the darkest greyscale
  corresponds to the peak intensity in the image.} 
\label{global2}
\end{center}
\end{figure*}

In order to search for emission from either the core of the lens galaxy
or a third image, we have also made maps of the region in between
images A and B. No evidence is found for either. If we assume that
these components would have been detected if their peak surface
brightness was greater than the 6-$\sigma$ rms noise level of
180~$\mu$Jy~beam$^{-1}$, we conclude that the third image has a flux
density less than 180~$\mu$Jy, 0.02~per~cent of the component A total
flux density. Here we have assumed that the third image will be
unresolved as its size should be that of A ($\sim 10 \times 10$~mas)
divided by the relative de-magnification factor of $\sim$5000 i.e. less
than the size of the beam. Multi-path scattering (which we show is
probably affecting the size of image A in Section~\ref{scattering})
could though broaden the third image and reduce its surface
brightness. If this is the case then the upper limit given above for
the flux density of the third image is an under-estimate.

\subsection{Global VLBI maps (polarised intensity)}
\label{polsec}

Our polarised intensity maps ($P=\sqrt{Q^{2}+U^{2}}$, bottom row of
Fig.~\ref{global1}) show no polarised jet emission outside of the
core region. Sub-components 1 and 2 though are prominent and in image A
are, as in the total intensity maps, tangentially stretched. In image
B polarised inter-component emission between B1 and B2 is visible and
appears as a small jet-like extension from sub-component B1, again
supporting the hypothesis that this is the core.

The top row of Fig.~\ref{global2} shows polarisation E-field vectors
overlaid on total intensity contours. The effect of the differential
Faraday rotation between the images can be seen as a rotation of the
polarisation vectors in image A by $\sim 50^{\circ}$ clockwise relative
to image B. In image B it is possible to identify
changes in the polarisation position angle along the jet axis,
including across B2, the least compact sub-component. Generally though
the position angle tends to lie approximately parallel to the jet
axis. This ``alignment'' is entirely coincidental as both the jet axis
and polarisation position angles are changed from their true values, the
first by the gravitational effect of the lens and the second by Faraday
rotation. In image A the distribution of the position angle is more 
complicated, there being an anomalous region of polarisation that
stretches northwards from the midway point between A1 and A2. Here the
polarisation vectors are oriented at $\sim 0^{\circ}$ as opposed to
elsewhere in this image where the majority of the polarisation is
oriented between position angles of $\sim$30--40$^{\circ}$. This
anomalous region is also seen in the maps of \citet{kemball01}.

The bottom row of Fig.~\ref{global2} shows (in greyscale) percentage
polarisation maps of A and B, plotted on the same angular scale as the
polarisation position angle figures above. The inner jet of B
is particularly prominent here as it is very highly polarised, with a
peak of 20~per~cent. Interestingly, the distribution of percentage
polarisation looks very different in image A where the peak
polarisation is lower at only 12~per~cent. There is also no sign of a
significantly more highly polarised sub-component between A1 and A2.

\subsection{Global VLBI maps (model fitting)}

We measured the flux densities and sizes for sub-components 1 and 2 by
fitting elliptical Gaussians to the total intensity maps using the {\sc
aips} tasks {\sc jmfit} and {\sc imfit}. We choose to fit to the maps,
rather than to the $(u,v)$ data, as the biasing effect of the extended
jet emission can then be more easily reduced. This is achieved by
removing all pixels from the images whose surface brightness is less
than 2.88~mJy~beam$^{-1}$. This corresponds to the 
sixth-lowest positive contour in the total intensity maps of
Fig.~\ref{global1}, above which there is little evidence of jet
emission in either image A or B. The possibility of contamination from
the inner jet remains and the effect of this, although probably small
due to this being relatively weak, will be different in each image. The
positions of the Gaussians are constrained to lie at the maxima of
sub-components 1 and 2.

\begin{table}
\begin{center}
\caption{Results of elliptical Gaussian fitting to the VLBI
  maps. Tabulated are flux densities ($\mathrm{S}_{\nu}$), deconvolved 
  major ($a$) and minor ($b$) axes and orientation of ellipticity
  ($\phi$).}
\begin{tabular}{llllr} \hline
Image & $\mathrm{S}_{\nu}$ (mJy) & $a$ (mas) & $b$ (mas) & $\phi$ ($^{\circ}$)\\ \hline
A1 & $270\pm5$ & $1.3\pm0.1$   & $0.5\pm0.1$   & $-33\pm2$ \\
A2 & $280\pm5$ & $2.3\pm0.1$   & $1.2\pm0.1$   & $-30\pm2$ \\ \hline
B1 & $114\pm3$ & $0.30\pm0.02$ & $0.16\pm0.02$ & $83\pm2$  \\
B2 & $68\pm3$  & $0.80\pm0.02$ & $0.40\pm0.02$ & $81\pm2$  \\
\end{tabular}
\label{modelfit}
\end{center}
\end{table}

The measured flux densities and deconvolved sub-component sizes are
shown in Table~\ref{modelfit}. Errors have been estimated by
comparing the results from {\sc imfit} and {\sc jmfit}, but are
probably underestimates of the true uncertainties due to the
systematic effects alluded to in the previous paragraph. If we consider
only image B, where the errors are smaller, we see that sub-component 1
is the more compact and that its flux density is significantly lower
than that measured by \citet{kemball01}. As that of B2 has remained
unchanged over the intervening five years, these maps support the
identification of sub-component 1 as the core and the origin of the
variability seen in the VLA monitoring campaigns.

\subsection{VLA maps (outer subarray)}
\label{vlasec}

\begin{figure}
\begin{center}
\includegraphics[scale=0.4]{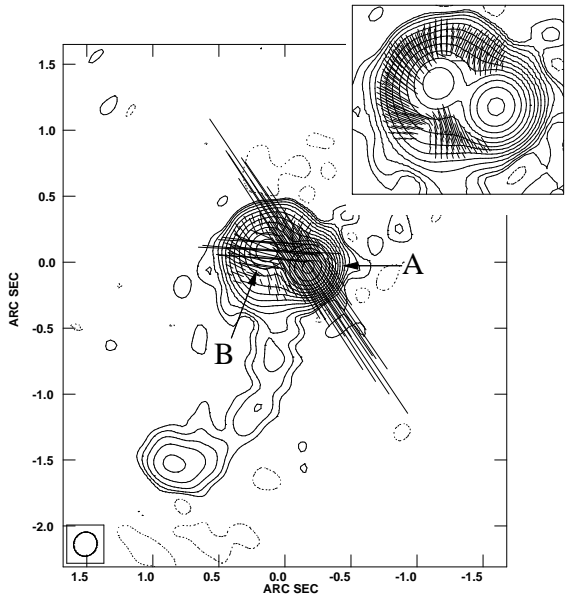}
\caption{VLA (outer subarray) contour map of JVAS B0218+357 at 8.4~GHz
  with polarisation E-vectors overlaid. The restoring beam is shown in
  the bottom-left corner of the map and has a FWHM of $186 \times
  174$~mas at a position angle 
  of $-26$\fdg2. Contours are plotted at multiples ($-1$, 1, 2, 4, 8, 16,
  etc) of 3$\sigma$ where $\sigma$ is the off-source rms noise in the map
  ($50~\mu$Jy~beam$^{-1}$). The inset at the top right shows the
  region of the Einstein ring without the A and B polarisation vectors.} 
\label{vlamap}
\end{center}
\end{figure}

The total intensity map of the VLA-only data is shown in
Fig.~\ref{vlamap} with polarisation E-field vectors overlaid. It shows
all the major features of B0218+357 on arcsec scales -- the radio cores
A and B, the Einstein ring and the radio jet. In addition, the region
around the Einstein ring is shown with the polarisation vectors
corresponding to images A and B omitted. This 
enables the polarisation vectors in the ring to be seen more clearly.
It is interesting to see that around the majority of the ring the
polarisation vectors are aligned radially outwards. As gravitational
lensing preserves the orientation of polarisation on the sky
\citep{dyer92} this is somewhat unexpected. Although it is possible
that the polarisation is Faraday rotated in the jet as well as in the
cores, it would be surprising if the spatial distribution of Faraday
depths was such that the polarisation became aligned with the total
intensity structure of the Einstein ring. However, this seems to be the
case.

\begin{table}
\begin{center}
\caption{Total flux density ($I$), polarised intensity, ($P$),
  percentage polarisation ($m$) and polarisation position angle ($\phi$)
  for images A and B as measured from the VLBI and VLA datasets. The
  $P$ and $\phi$ VLBI values are derived from measurements of the
  integrated $Q$ and $U$ flux densities. The VLA values have been
  calculated by model-fitting delta components to the $(u,v)$ data.}
\begin{tabular}{lllll} \hline
Image & $I$ (mJy) & $P$ (mJy) & $m$ (\%)& $\phi$ ($^{\circ}$)\\ \hline
A (VLBI) & 602 & 35.8 & 5.9 & 29 \\
A (VLA)  & 744 & 34.1 & 4.6 & 33 \\
B (VLBI) & 191 & 17.1 & 9.0 & 83 \\
B (VLA)  & 253 & 16.8 & 6.6 & 82 \\
\end{tabular}
\label{modtab}
\end{center}
\end{table}

We have estimated the flux densities of images A and B by fitting two
delta components to the $(u,v)$ data in {\sc difmap}
\citep{shepherd97}. The results are shown in Table~\ref{modtab}. In
fitting to the Stokes $I$ data we include a third component (a ring,
component type 4 of the {\sc difmap} modelfit function) to approximate
the bright Einstein ring structure and 
improve the fit. No such additional component was added when fitting to
the Stokes $Q$ and $U$ maps as the Einstein ring is much fainter. Also
included in Table~\ref{modtab} are values for the total intensities,
polarised intensities and polarisation position angles as measured from
the VLBI data. These are calculated from the integrated $I$, $Q$
and $U$ flux densities which facilitates a comparison with the values
measured from the lower-resolution VLA (outer subarray) dataset. From
the $P$ and $I$ measurements we calculate the corresponding percentage
polarisations, $m$ (\%), which are also tabulated.

The good correspondence between the polarisation position
angles as measured from the VLA and VLBI data suggests that the
absolute position angle calibration has been successful. Similarly, the
excellent correspondence between the integrated polarised intensities
signifies that the flux scale of the two arrays are correctly aligned.
However, in the VLBI maps we only detect 81~per~cent of the VLA flux
density in image A 
and 75~per~cent in image B. We think this is due to a combination of
the poorer surface-brightness sensitivity of the VLBI array (most of
the jet visible in Fig.~\ref{cj} is not detected) and over-estimation
of the VLA flux densities due to blending with the ring emission in the
map. This latter hypothesis is supported by the fact that we detect
more of the VLA flux density in image A (which 
is less blended with the ring in the VLA map) as well as by the very
similar polarised intensities measured by the VLA and VLBI arrays. We
have made tapered maps of the VLBI data in an attempt to recover more
of the A and B image flux densities. More is detected ($\sim$5~per~cent
in each image), but the VLBI flux densities of A and B still fall
significantly short of their VLA values.

\section{Discussion}

Our global VLBI 8.4-GHz maps have revealed a great deal of substructure
in images A and B of JVAS B0218+357. Neither the lens galaxy core nor
the third image were detected, both of which would be very useful
constraints on the lens model, especially in this system where the
centre of the lensing galaxy is currently uncertain.\footnote{The lack
  of detection of a third image can be used though to place an upper
  limit on a finite core radius of the lens galaxy. This has been done
  for B0218+357 by \citet{norbury01}.} In 
this section we discuss three main topics: the constraints that the new
maps can already put on lens mass models, the evidence that the image
suffers from multi-path scattering and the different polarisation
distributions in each image.

\subsection{Mass model constraints}

\begin{figure*}
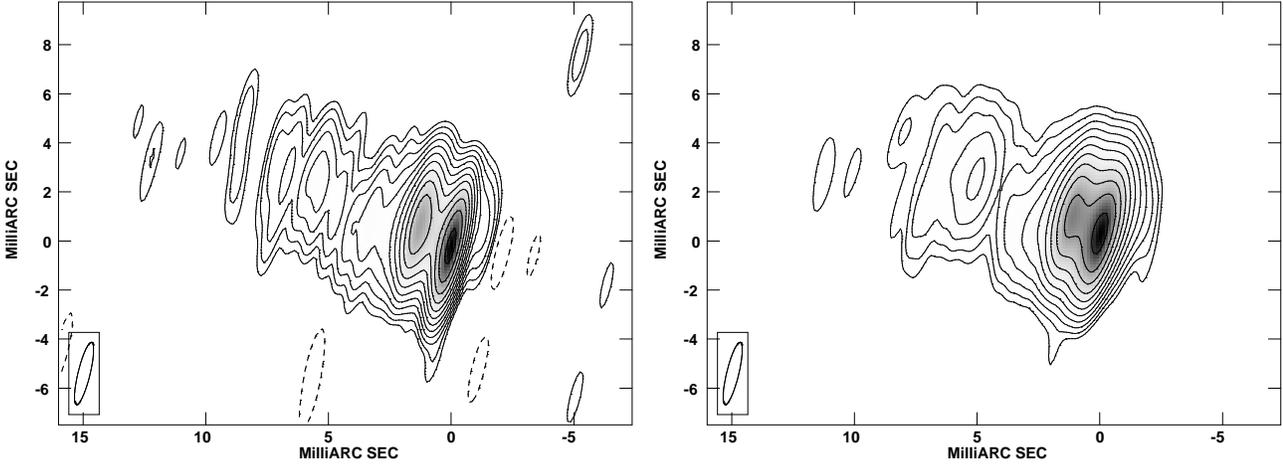

\begin{center}
\includegraphics[scale=0.45]{fig5b.ps}
\includegraphics[scale=0.45]{fig5a.ps}
\caption{Maps of images A (right) and B (left) after being
  back-projected into the source plane. The restoring beam is shown in the
  bottom-left corner and has a FWHM of $2.61 \times 0.49$~mas 
  at a position angle of $-13$\fdg7.} 
\label{backproj}
\end{center}
\end{figure*}

The extensive substructure that we have detected will be useful for lens
modelling and ultimately it is our intention to use the LensClean
algorithm on these data for this purpose, the results of which will be
presented in a future paper. In the interim we have back-projected each
image into the source plane in order to qualitatively demonstrate the 
accuracy of the model. This has been done using the {\sc clean}
components and an isothermal model of the lens galaxy optimized using
the positions of the 15GHz VLBI sub-components \citep{patnaik95}, a flux
density ratio of 3.75 and the galaxy position found from our LensClean
of VLA and MERLIN data \citep{wucknitz01}. With $x$ and $y$ pointing
along the major and minor axis of the ellipsoid, the potential is
written as 
\begin{equation}
\label{thefirstequation}
\phi= \alpha_0 \sqrt{\frac{x^2}{(1+\epsilon)^2} +
\frac{y^2}{(1-\epsilon)^2}}
\quad .
\end{equation}
The critical radius is $\alpha_0=160.6\,\rmn{mas}$, the ellipticity
$\epsilon=0.0583$ and the position angle of the major axis
$\theta=-41.9\,\rmn{deg}$, measured from north through east.  The centre
of the lens is at $\bmath{z}_0 = (260,117.5)\,\rmn{mas}$ relative to
A.

In order to compensate for the very different effective resolution of A
compared to B, the back-projected CLEAN components of both images have
been restored with the same beam. This beam is just large enough that
both the A and B back-projected beams are encompassed by it, which in
practice means that it is approximately equal to the restoring beam in
the image plane back-projected from image B to the source plane. This
makes the nominal resolution in each the same. The back-projected maps
are shown in Fig.~\ref{backproj}.

Compared to the lens-plane images of Fig.~\ref{global1}, the
source-plane images look quite similar. Worthy of note is the excellent
correspondence between the counter-components to the west of the core
region which are probably images of a weak counterjet. This is
intriguing though as the lensed radio source in B0218+357 has been
classified as a BL~Lac \citep{odea92,browne93,stickel93}, a class of
objects that are not expected to display counterjets according to the
unification schemes of extragalactic radio sources. This theoretical
prediction is supported by the non-detection of counterjets in BL~Lac
sources, a notable exception being PKS~1413+135
\citep{perlman94,perlman96}.

A closer comparison of the A and B source-plane maps shows that the B
jet seems to be stretched by about 10~per~cent relative to the A jet
i.e. the jet in B is longer. This stretching can only be explained with
a different radial mass profile, compared to that used in the model,
since the jet is directed more or less radially relative to the
galaxy's centre. Writing the radial surface mass density as a function
of radius as a power law, $\Sigma(r) \propto r^{-\beta}$, we find
$\beta\approx1.04$. A very similar value is found ($\beta=1.06\pm0.03$)
if we use the 15-GHz sub-component positions of \citet{patnaik95} as
constraints \citep{wucknitz01}.

\subsection{Multi-path scattering}
\label{scattering}
In Fig.~\ref{backproj} image A looks smoother than image B. We suggest
that this is due to scatter-broadening, which may also explain the
frequency-dependent sizes of the images at low ($\le$2.3~GHz)
frequencies (Biggs et al., in preparation). Other examples of where
scattering is believed to modify the surface brightness of
gravitationally lensed images are PKS~1830-211
\citep{jones96,guirado99} and CLASS B1933+503 \citep{marlow99}. In
B0218+357, as well as in the other two systems, the scattering would
most likely originate in the lensing galaxy as we know from the observed
Faraday rotation and differential depolarisation that this is both
highly ionised and non-homogeneous. Also, at the lens redshift of
$z=0.6847$ it is perfectly reasonable to expect that the ISM in front
of each image, 
with a separation of 334~mas (2.4~kpc)\footnote{Throughout this paper
we assume a flat universe with $H_0=70$\,km\,s$^{-1}\,$Mpc$^{-1}$, 
$\Omega_0=0.3$ and $\lambda_0 = 0.7$.} between them, could be
significantly different.

In order to quantify the effect, we have fitted Gaussians to
sub-components 1 and 2 of the back-projected images of A and B using
the task {\sc jmfit}. The deconvolved major ($a$) and minor ($b$) axes,
as well as the position angle of the major axis ($\phi$), are shown in
Table~\ref{jmfit} where it can be seen that for each sub-component the
geometric mean of the axes (equivalent circular size) is bigger in
A than in B; the area of the deconvolved A1 is approximately eight
times that of B1. Also, whilst the position angles of B are as expected
(they point more or less down the jet as in the image plane) those of A
remain approximately perpendicular to this.

The advantage of model-fitting to the core sub-components in the source
plane is that the back-projection removes the differences between the
two images, reducing the need for removal of confusing pixels
corresponding to the jet emission and rendering any bias that might
exist identical in each. The disadvantage of fitting in the source
plane is the potential for errors to accumulate during the many
convolutions and deconvolutions that have been applied to the data at
various stages of the data reduction. However, we note that the
measured sizes in the image plane also show that image A is larger than
image B, by a factor greater than the relative magnification. From the
deconvolved sizes of Table~\ref{modelfit} the image-plane area ratios
A1/B1 and A2/B2 are approximately 14 and 9. For the remainder of this
analysis we only consider the source-plane size measurements.

\begin{table}
\begin{center}
\caption{Deconvolved major ($a$) and minor ($b$) axes and orientation
  of ellipticity ($\phi$) of Gaussians fitted to the back-projected
  maps as well as their geometric means and the size of the derived
  scattering discs ($\theta_{\mathrm{sc}}$). All quantities are
  measured in mas apart from position angles which are measured in
  degrees. We estimate errors on the major and minor axes of about
  5~per~cent.}
\begin{tabular}{lllrll} \hline
Image & $a$ & $b$ & $\phi$ & $\sqrt{a\times b}$ & \multirow{1}*{$\theta_{\mathrm{sc}}$} \\ \hline
A1 & 1.07 & 0.67 & 151 & 0.85 & \multirow{2}*{0.79} \\
B1 & 0.34 & 0.28 & 57  & 0.31 & \\ \hline
A2 & 1.24 & 0.96 & 149 & 1.09 & \multirow{2}*{0.83} \\
B2 & 0.78 & 0.64 & 76  & 0.71 & \\
\end{tabular}
\label{jmfit}
\end{center}
\end{table}

We assume that the observed size of an A sub-component is the sum,
in quadrature, of the intrinsic size and a scattering scale size
($\theta_{\mathrm{sc}}$). We further assume that, for each
sub-component, the true intrinsic size is actually that of its image B
counterpart i.e. that this image is unaffected by scattering at this
frequency. This assumption is supported by the similarity of the
deconvolved size of B1 as measured from these data and the 15-GHz data
of \citet{patnaik95}. The derived size of $\theta_{\mathrm{sc}}$ is
also shown in Table~\ref{jmfit} and is the same for both
sub-components, 0.8~mas. The true size of the scattering disc (we have
measured sub-component sizes in the source plane) is
\begin{equation}
\label{thesecondequation}
\hat{\theta}_{\mathrm{sc}}=\theta_{\mathrm{sc}}
\frac{D_{\mathrm{s}}}{D_{\mathrm{ds}}},
\end{equation}
where $D_{\mathrm{s}}$ is the angular diameter distance between the
Earth and the lensed source and $D_{\mathrm{ds}}$ that between the
lensing galaxy and the lensed source. The relation in
equation~(\ref{thesecondequation}) is analogous to that between the true
deflection angle ($\hat\alpha$) and the {\em apparent} deflection angle
($\alpha$) \citep[e.g.][]{refsdal94}. No correction for the lens
magnification is required as the displacement in the image plane due to
scattering is the scattering angle magnified by the lens, and this was 
removed when back-projecting each image. We calculate
$\hat{\theta}_{\mathrm{sc}} = 3.4$~mas.

The size of the scattering disk is equal to the refractive length scale
($r_{\mathrm{ref}}$) subtended at the distance of the lensing galaxy
\citep{narayan92}. Using the formulation of \citet{walker98} we have
estimated the scattering strength (SM) of the ionised medium. With the
observed frequency of 8.4~GHz corrected to the redshift of the lensing
galaxy we measure $\mathrm{SM \approx 150~kpc~m^{-20/3}}$. This is a
very large value compared to those seen along typical lines of sight
through the Galaxy, although comparable (and in some cases much higher)
values have been measured, particularly towards the Galactic centre
\citep{cordes02}. Therefore, although the scattering in front of
B0218+357 image A is extreme, it is not ruled out by observations in
the Galaxy.

\subsection{Depolarisation}

The polarisation properties of this system continue to intrigue. From
observations of this system with MERLIN and the VLA, a general picture
has emerged of image A being less polarised than image B and of both
depolarising with increasing wavelength, albeit much more steeply
in A. The reason for this is probably inhomogeneities in the
magneto-ionic medium that is responsible for the Faraday rotation of the
polarisation position angles, inhomogeneities that are 
greater in the region of A than B. Complications that hinder the
interpretation of the polarisation results include ``beating'' of the
polarisation position angles of sub-components 1 and 2 (for
observations which do not resolve the sub-components), changing source
structure with frequency and different magnification gradients across
each image.

The resolution afforded by VLBI observations allows the polarisation
structure of the source to be resolved and constraints to be put on the
angular scale of the inhomogeneities in the Faraday screen. In
Fig.~\ref{global2} we see that image B, the least resolved of the
images, is seen to have a higher peak polarisation (20~per~cent) than
that of A (12~per~cent). This shows that we are not seeing the effects
of {\em intrinsic} changes in the polarisation structures across the
radio source as the lower effective resolution of B would cause these
to be averaged incoherently more than in A i.e. we would expect image B
to be less polarised than image A. Therefore, our global VLBI maps 
support the theory that the depolarisation is due to non-uniformities
in the Faraday screen which are greater in front of image A.
They also allow us to constrain the size of these irregularities
which must be smaller than the synthesised beam i.e. $\la$1~mas. This
corresponds to a linear distance of $\la$7~pc at the redshift of the
lensing galaxy.

We also note that the scatter-broadening identified in
Section~\ref{scattering} could cause the depolarisation. This is
because scatter-broadening effectively blurs a source, causing regions
with different polarisations to overlap and be averaged together. As
the scattering is different in front of each image the polarisations of
each will be different and, as scattering 
increases at longer wavelengths, the observed frequency dependence of
the depolarisation could also result.

Another interpretation is that the differences between the images could
result from the combination of intrinsic polarisation 
variability and time delay. VLA monitoring \citep{corbett96,biggs99}
has shown that this source is variable in polarisation as well as in
total flux density. Two-epoch VLBI observations separated by the time
delay, $10.5\pm0.4$~d, could help to clarify the polarisation structure
in this source. At the same time the picture of A being more
depolarised than B is too simplistic. From earlier VLA monitoring 
observations \citep{corbett96} it was found that, with the time delay
removed, the polarisation of image B was {\em lower} than that of image
A at 15~GHz, the opposite to what is usually observed. This could
signify time variability within the magneto-ionic medium or movement of
the source relative to this.

\section{Summary and future work}

We have presented new maps of the lens system JVAS~B0218+357 made with
a global VLBI array that combine excellent sensitivity and resolution
to produce the best maps yet of the mas-scale structure of this
system. Unfortunately, we were unsuccessful in detecting the
third image or the core of the lensing galaxy. With the new maps we
have been able to explore several aspects of the lensing galaxy. From
the different lengths of the jet in each image we have been able to
constrain the radial mass profile and find that it is slightly steeper
than isothermal ($\beta\approx1.04$). We also find evidence for
multi-path scattering and depolarisation due to the disparity between
the component sizes and polarisations in each image. These physical
effects most likely result from propagation through the turbulent
ionised ISM of the lensing galaxy and, in this way, B0218+357 is
probably the best example of how gravitational lensing can act as a
unique probe of the ISM of high-redshift galaxies.

These data will be used with LensClean \citep{kochanek92} in order to
exploit fully the resolved structure of the images and provide further
constraints for the lens model. Whilst the effects of
scatter-broadening might seem to limit the usefulness of these data for
lens modelling, we note that the effects of the scattering and the lens
can be separated. This is because scattering only effects
the sizes of components whilst the lens changes both the component
sizes and separations. Therefore, future modelling will be able to not only
constrain the lens model, but also to characterise the scattering more
accurately than has been possible in this paper.

Further observations with a Global VLBI array at 5~GHz may be
worthwhile to try and detect, with higher dynamic range, the jet
emission that has been revealed in the new CJ maps (Fig.~\ref{cj}). The
present and future VLBI maps will be used to constrain the lens model,
in conjunction with MERLIN, VLA and $HST$ data. $HST$ observations will
be carried out with the Advanced Camera for Surveys which our
simulations predict will give us an accuracy of $\sim$10~mas in the
galaxy position relative to the lensed images. With this vital
parameter in the lens model finally measured directly (as opposed to
the {\em modelled} position found from LensClean) the uncertainty in
the $H_0$ determination from this system will be reduced to a level of
about 3~per~cent.

\section*{Acknowledgments}

The VLBA and VLA are operated by the National Radio Astronomy
Observatory which is a facility of the National Science Foundation
operated under cooperative agreement by Associated Universities,
Inc. The European VLBI Network is a joint facility of European and
Chinese radio astronomy institutes funded by their national research
councils. We would like to thank the staff of the DSN at Goldstone and
Robledo for their assistance in making these observations possible. This
research was supported in part by the European Commission TMR
Programme, Research Network Contract ERBFMRXCT96-0034 `CERES'. OW was
partially funded during this work by the European Commission, Marie
Curie Training Site programme, under contract
no. HPMT-CT-2000-00069. ADB would like to thank the staff of JIVE for  
their help with writing the observing schedule. Much thanks goes to
Alok Patnaik for his contribution to the proposal as well as for his
help with the initial stages of the data reduction. We also thank the
anonymous referee for a number of suggestions and comments that
improved the paper.

\end{document}